\documentclass[english,useAMS,usenatbib]{mn2e} 
\usepackage{lmodern}

\usepackage[T1]{fontenc}
\usepackage[utf8]{inputenc}
\usepackage{color}
\usepackage[usenames,dvipsnames]{xcolor}
\usepackage{amsmath}
\usepackage{amssymb}
\usepackage{graphicx}
\usepackage{esint}
\usepackage{placeins}
\usepackage[authoryear]{natbib}

\usepackage{hyperref}
\definecolor{darkgreen}{rgb}{0.1,0.6,0.7}
\hypersetup{colorlinks = true,urlcolor=blue,citecolor=Blue}

\makeatletter
\usepackage{lmodern}\usepackage{epstopdf}

\let\jnl@style=\rm
\def\ref@jnl#1{{\jnl@style#1}}

\def\aj{\ref@jnl{AJ}}                   
\def\actaa{\ref@jnl{Acta Astron.}}      
\def\araa{\ref@jnl{ARA\&A}}             
\def\apj{\ref@jnl{ApJ}}                 
\def\apjl{\ref@jnl{ApJ}}                
\def\apjs{\ref@jnl{ApJS}}               
\def\ao{\ref@jnl{Appl.~Opt.}}           
\def\apss{\ref@jnl{Ap\&SS}}             
\def\aap{\ref@jnl{A\&A}}                
\def\aapr{\ref@jnl{A\&A~Rev.}}          
\def\aaps{\ref@jnl{A\&AS}}              
\def\azh{\ref@jnl{AZh}}                 
\def\baas{\ref@jnl{BAAS}}               
\def\bac{\ref@jnl{Bull. astr. Inst. Czechosl.}}
\def\caa{\ref@jnl{Chinese Astron. Astrophys.}}
\def\cjaa{\ref@jnl{Chinese J. Astron. Astrophys.}}
\def\icarus{\ref@jnl{Icarus}}           
\def\jcap{\ref@jnl{J. Cosmology Astropart. Phys.}}
\def\jrasc{\ref@jnl{JRASC}}             
\def\memras{\ref@jnl{MmRAS}}            
\def\mnras{\ref@jnl{MNRAS}}             
\def\na{\ref@jnl{New A}}                
\def\nar{\ref@jnl{New A Rev.}}          
\def\pra{\ref@jnl{Phys.~Rev.~A}}        
\def\prb{\ref@jnl{Phys.~Rev.~B}}        
\def\prc{\ref@jnl{Phys.~Rev.~C}}        
\def\prd{\ref@jnl{Phys.~Rev.~D}}        
\def\pre{\ref@jnl{Phys.~Rev.~E}}        
\def\prl{\ref@jnl{Phys.~Rev.~Lett.}}    
\def\pasa{\ref@jnl{PASA}}               
\def\pasp{\ref@jnl{PASP}}               
\def\pasj{\ref@jnl{PASJ}}               
\def\rmxaa{\ref@jnl{Rev. Mexicana Astron. Astrofis.}}%
\def\qjras{\ref@jnl{QJRAS}}             
\def\skytel{\ref@jnl{S\&T}}             
\def\solphys{\ref@jnl{Sol.~Phys.}}      
\def\sovast{\ref@jnl{Soviet~Ast.}}      
\def\ssr{\ref@jnl{Space~Sci.~Rev.}}     
\def\zap{\ref@jnl{ZAp}}                 
\def\nat{\ref@jnl{Nature}}              
\def\iaucirc{\ref@jnl{IAU~Circ.}}       
\def\aplett{\ref@jnl{Astrophys.~Lett.}} 
\def\apspr{\ref@jnl{Astrophys.~Space~Phys.~Res.}}
\def\bain{\ref@jnl{Bull.~Astron.~Inst.~Netherlands}} 
\def\fcp{\ref@jnl{Fund.~Cosmic~Phys.}}  
\def\gca{\ref@jnl{Geochim.~Cosmochim.~Acta}}   
\def\grl{\ref@jnl{Geophys.~Res.~Lett.}} 
\def\jcp{\ref@jnl{J.~Chem.~Phys.}}      
\def\jgr{\ref@jnl{J.~Geophys.~Res.}}    
\def\jqsrt{\ref@jnl{J.~Quant.~Spec.~Radiat.~Transf.}}
\def\memsai{\ref@jnl{Mem.~Soc.~Astron.~Italiana}}
\def\nphysa{\ref@jnl{Nucl.~Phys.~A}}   
\def\physrep{\ref@jnl{Phys.~Rep.}}   
\def\physscr{\ref@jnl{Phys.~Scr}}   
\def\planss{\ref@jnl{Planet.~Space~Sci.}}   
\def\procspie{\ref@jnl{Proc.~SPIE}}   

\newcommand{\half}{\frac{1}{2}}

\newcommand{\calW}{\ensuremath{\mathcal{W}}}

\newcommand{\boSig}{\ensuremath{{\sf{\Sigma}}}}

\newcommand{\mathd}{\ensuremath{\mathrm{d}}}

\newcommand{\both}{\ensuremath{\boldsymbol{\theta}}}

\newcommand{\calP}{\ensuremath{\mathcal{P}}}

\newcommand{\calG}{\ensuremath{\mathcal{G}}}

\newcommand{\fatx}{\ensuremath{\boldsymbol{x}}}

\newcommand{\sfC}{\ensuremath{{\sf{C}}}}
\newcommand{\sfG}{\ensuremath{{\sf{G}}}}

\newcommand{\GSN}{\ensuremath{{\sfG^{\rm SN}_\ell}}}
\newcommand{\GCV}{\ensuremath{{\sfG^{\rm CV}_\ell}}}

\newcommand{\sfX}{\ensuremath{{\sf{X}}}}

\newcommand{\fsky}{\ensuremath{f_{\mathrm{sky}}}}

\title[Euclid-era cosmology for everyone]{Euclid-era cosmology for everyone: Neural net assisted MCMC sampling for the joint 3x2 likelihood}

\author[Manrique-Yus \& Sellentin]{Andrea Manrique-Yus$^{1}$, Elena Sellentin$^{1}$\\
$^{1}$Leiden Observatory, Leiden University, Huygens Laboratory, Niels Bohrweg 2, NL-2333 CA Leiden, The Netherlands.
}

\makeatother

\usepackage{babel}
\begin{document}
\setlength{\voffset}{-12mm} 

\date{Accepted: 3rd Oct 1990. Received: 9th Nov 1989; in original form: Mondays, Fall 1989.}

\maketitle
\pagerange{\pageref{firstpage}--\pageref{lastpage}} \pubyear{2016}

\label{firstpage} 
\begin{abstract}
We develop a fully non-invasive use of machine learning in order to enable open research on Euclid-sized data sets. Our algorithm leaves complete control over theory and data analysis, unlike many black-box like uses of machine learning. Focusing on a `3x2 analysis' which combines cosmic shear, galaxy clustering and tangential shear at a Euclid-like sky coverage, we arrange a total of 348000 data points into data matrices whose structure permits not only an easy prediction by neural nets, but it additionally permits the essential removal from the data of patterns which the neural nets could not `understand'. The latter provides an often lacking mechanism to control and debias the inference of physics. 
The theoretical backbone to our neural net training can be any conventional (deterministic) theory code, where we chose CLASS. After training, we infer the seven parameters of a $w$CDM cosmology by Monte Carlo Markov sampling posteriors at Euclid-like precision within a day. We publicly provide the neural nets which memorise and output all 3x2 power spectra at a Euclid-like sky coverage and redshift binning.
\end{abstract}


\section{Introduction}
The advent of ever larger cosmic surveys requires ever faster numerical methods to analyse their data. With Planck \citep{Planck2015,Planck2018} having proven enormously successful in constraining not only the cosmological standard model, but also many non-standard models through extensive re-analyses by the community, one would like to enable such re-analyses also for Euclid-sized data sets \citep{Euclid}, or equally reanalyses of the Large Synoptic Sky Survey (LSST), NASA's WFIRST, or the Square Kilometer Array (SKA) \citep{Synergies,LSSTScienceBook,SKA}. Ultimately, a fusion of data sets from the early Universe (Planck) and from the late Universe will enable the physics of the Universe to be probed throughout its history, but this again constitutes a formidable computational challenge.

Ideally though, numerical challenges should not be felt by the community. Consequently, we here enable theoretical predictions from highly accurate (but slow) codes which compute physics beyond the standard model. In fact, while designing likelihoods for Euclid-sized weak lensing surveys \citep{SHH18, SH15}, we found that the computational lion's share in likelihood evaluation is solely due to computing model predictions from theoretical physics. The same bottleneck has been reported by \cite{EuclidEmulator}. We therefore consider it desirable to have a method in place, which runs automatically in the background of any likelihood, and accelerates the theory computations, no matter which theory code is plugged in. Providing such a method is the aim of this paper. A complementary approach has been developed for Einstein-Boltzmann solvers by \cite{Albers}, where neural nets are used to replace the expensive integration of differential equations.

In this paper, we base our method on a fully non-invasive use of artificial neural nets. We define non-invasive to mean that the cosmological inference does not \emph{depend} on the black-box like inner workings of a neural net: the neural net could always be switched off, and computing the likelihood would then require a (much) larger computing cluster, but still proceed along precisely the same path. As not all institutes will have a cluster that matches their theoretical models' numerical complexity in the Euclid-era, the neural net will simply decrease the numerical power needed.

The main problem in using a neural net in physical inference is of course that a neural net is by construction a universal approximator: with the approximation comes per se a loss of accuracy, which is a problem that must be addressed. In this paper, we shall solve it by `cleaning the data', i.e. removing from the data those fine structures that were not `understood' by the neural net. Omitting this step would bias the inference to an unknown degree, which is in fact one of the most often voiced caveats against use of machine learning in physics.

In section~\ref{sec:likelihood} we describe our setup of a joint 3x2 likelihood for Euclid. This is to be understood as a forecasting-like setup with the same numerical complexity as the upcoming real likelihoods. In section~\ref{sec:NN} we discuss what the neural nets `learn', and why this still leaves full control to theoretical physicists over their theory. Finally, section~\ref{sec:MCMC} shows the results from Monte Carlo Markov Chain (MCMC) sampling, and section~\ref{sec:conclusions} concludes our paper.

A further advantage of our method is that alongside any \emph{data} release, pre-computed theoretical predictions can also be publicly released, as a trained neural net constitutes a query-able memory. We therefore provide our public code and our trained `memories' of theory computations at \href{https://github.com/elenasellentin/CosmicMemory}{https://github.com/elenasellentin/CosmicMemory}. The physics memorised by our public neural net is a $w$CDM model, which uses cold dark matter (CDM), and two equation of state parameters for dark energy. For the special values $w_0 = -1$ and $w_a = 0$ of the equation of state parameters, the model (and the neural net) produce $\Lambda$CDM predictions, where $\Lambda$ is the cosmological constant.

\begin{table}
\caption{Adopted fiducial cosmology and priors. The fiducial cosmology is used to create a synthetic data set, and the priors are multiplied to the likelihood when converting to a posterior.}
\label{tab:example}
\begin{tabular}{lccr}
\hline
parameter & fiducial value & prior shape & prior bounds\\
\hline
$\Omega_{\rm cdm}$        & 0.315 & flat & [0.2,0.4] \\
$\Omega_{\rm DE}$ & $1-\sum_i \Omega_i$ & NA & NA \\
$\Omega_{\rm b}$  & 0.049 & BBN, flat & [0.02,0.06] \\
$h$               & 0.7 & flat &  [0.5,0.9] \\
$\sigma_8$        & 0.811 & flat & [0.65,0.95] \\
$n_s$             & 0.965 & flat & [0.9,1.0] \\
$w_0$             & -1.0 & flat & [-1.5,-0.66] \\
$w_a$             & 0.0 & flat & [-1,1] \\
\hline
\end{tabular}
\end{table}

\section{Setup of data vector and likelihood}
\label{sec:likelihood}
We work on the celestial sphere, for the dual reason of noise and beyond-$\Lambda$CDM theories being more accurately treatable on the sphere: beyond-$\Lambda$CDM theories typically affect the largest scales, where sky curvature is non-negligible \citep{Rel1,Rel2,Rel3,Rel5,Rel6}, and due to the sphere being compact, most statistical calculations simplify.

Our full-sky likelihood follows \cite{Hamimeche1,Hamimeche2} and \cite{SHH18}, which describe likelihoods for power spectra of spherical harmonics. The estimated power spectra are compared to sets of theoretical predictions $\{ C_\ell(\both) \}$ of these power spectra, which we will describe below.

We denote a unit vector indicating the direction on the sphere as $\vec{n}$, and expand an observed field $\Phi(\vec{n})$ in spherical harmonics $Y_{\ell m}$, such that at celestial position $\vec{n}$ we have
\begin{equation}
    \Phi(\vec{n}) = \sum_{\ell,m} a_{\ell m}^\Phi Y^{(s)}_{\ell m}(\vec{n}),
\end{equation}
where $s$ is a potential spin-weight. The indices $\ell,m$ denote $\ell$-modes and $m$-modes respectively.
For two different fields $\Phi(\vec{n})$ and $\Psi(\vec{n})$, there will exist auto-power spectra ($\Phi = \Psi$) and cross-power spectra ($\Phi \neq \Psi$), which can be estimated by averaging over $m$-modes
\begin{equation}
    C^{\Phi, \Psi}(\ell) = \frac{1}{\nu} \sum_{m = - \ell}^\ell a_{\ell m}^{\Phi} (a_{\ell m}^\Psi)^*, 
    \label{estimators}
\end{equation}
where the asterisk denotes complex conjugation. We use the degrees of freedom
\begin{equation}
    \nu = \fsky (2\ell +1),
    \label{nu}
\end{equation}
where $\fsky$ denotes the sky fraction of the survey.

Theoretical cosmology can predict the expectation values of these power spectra. Denoting expectation values by angular brackets, we have
\begin{equation}
\bar{C}^{\Phi,\Psi}(\ell) = \langle C^{\Phi,\Psi}(\ell)\rangle,
\end{equation}
where the overbar indicates that these are the theoretical predictions.

\begin{figure}
\includegraphics[width=0.45\textwidth]{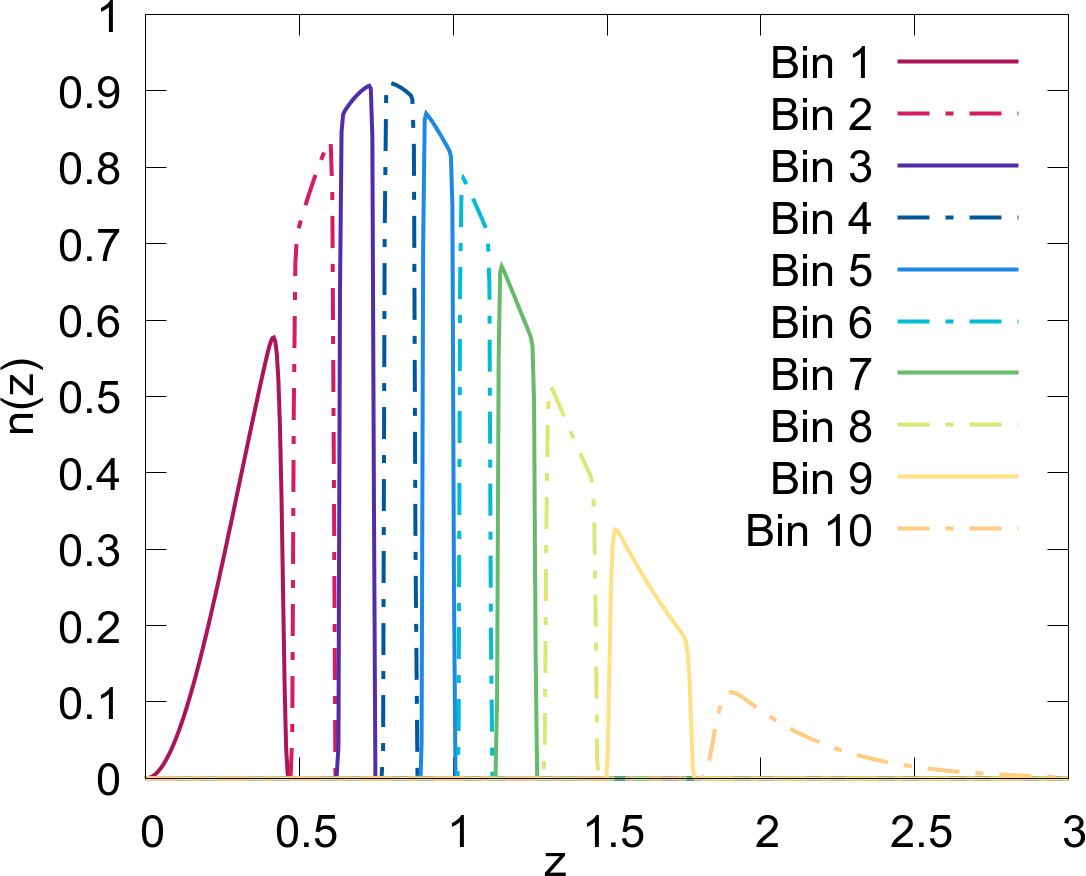} 
\caption{Setup of our tomographic redshift bins for the Euclid-like survey. Means and width for the bins are given in Table~\ref{tab:red}.}
\label{Redbins}
\end{figure}

For a Euclid-like observation of weak lensing, galaxy clustering and their cross correlation, the power spectra are usually arranged into a data vector, but as they are (co)variances of the underlying $a_{\ell m}$-modes, we shall here arrange them into data \emph{matrices} which correspond to the covariance matrices of the observed cosmic structures.

There will exist a data matrix per $\ell$-mode, of a somewhat rich structure, due to the three probes and the tomographic binning in redshifts. We denote a spherical harmonics power spectrum as $C_{z_1,z_2}^{\Phi,\Psi}(\ell)$, where the two lower indices denote the redshifts bins, and the two upper indices indicate the observed fields. We denote the field of galaxy overdensities as $g$, and the lensing potential as $\psi$. For a joint analysis of galaxy clustering and weak lensing, we then have to homogenise the spin-weights of our fields as follows, as the data matrix could otherwise not be a proper covariance matrix of multiple fields with different spin-weights.

Galaxy clustering $g$ and the lensing potential $\psi$ are both scalar fields, and hence spin-0. The observable most easily extractable from weak lensing is however the shear $\gamma$, which is spin-2. There can however not be a cross-correlation between a spin-0 and a spin-2 field, hence we imagine that shears $\gamma$ are measured, of which the spherical harmonic power spectrum is then
\begin{equation}
    C^{\gamma,\gamma}_{ij} (\ell) = \frac{1}{4} \frac{(\ell+s)!}{(\ell-s)!} C^{\psi,\psi}_{ij}(\ell),
\end{equation}
where $C^{\psi,\psi}_{ij}(\ell)$ is the lensing potential power spectrum. We thus go via the observable spin-2 shear to spin-0 lensing potential, and then to convergence for the cross-correlation.  Given an estimated power spectrum $C^{\gamma, \gamma}_{ij}(\ell)$, the associated convergence power spectrum is
\begin{equation}
C^{\kappa,\kappa}_{ij}(\ell) = \frac{[\ell(\ell+1)]^2}{4} C_{ij}^{\psi,\psi}(\ell),
\end{equation}
where $\kappa$ is the convergence. 

\begin{figure}
\includegraphics[width=0.45\textwidth]{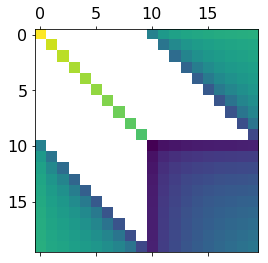} 
\caption{Plot of the logarithm of a Euclid-like 10-bin data matrix per $\ell$-mode. The upper diagonal block is galaxy clustering, where the zero off-diagonal elements arise from our non-overlapping bin definition. The lower blueish diagonal block is weak lensing, and the triangular off-diagonal plots are the absolute value of cross correlations between galaxy clustering and weak lensing. Each matrix element is a (cross) power spectrum at fixed $\ell$.}
\label{EuclidMatrixPlot}
\end{figure}

The cross-correlation with galaxy clustering can now be theoretically predicted and be used in a covariance matrix. Its associated estimator is `tangential shear', and the predicted associated cross power spectrum is then \citep{HuA11,Rel3,Kilb}
\begin{equation}
C^{g,\kappa}_{ij}(\ell) = - \frac{\ell(\ell+1)}{2} C_{ij}^{g,\psi}.
\end{equation}
The power spectra $C^{\kappa,\kappa}_{ij}$, $C^{g,g}_{ij}$ and $C^{g,\kappa}_{ij}$ can now be assembled into a sensible covariance matrix of the spherical harmonic coefficients $a_{\ell m}^\kappa$ and $a_{\ell m}^g$.
For a survey with three tomographic bins, this data matrix per $\ell$-mode is then the block-matrix  
\begin{equation}
\hat{\sfG}_\ell =
\arraycolsep=1.4pt\def\arraystretch{1.5}
\def\arraystretch{1.4}
\left( \begin{array}{ccc|ccc}
C^{g,g}_{1,1} & 0 & 0 & C^{g,\kappa}_{1,1} & C^{g,\kappa}_{1,2} & C^{g,\kappa}_{1,3}\\
& C^{g,g}_{2,2} & 0 & 0 & C^{g,\kappa}_{2,2} & C^{g,\kappa}_{2,3}\\
& & C^{g,g}_{3,3} & 0 & 0 & C^{g,\kappa}_{3,3} \\
\hline
& & & C^{\kappa,\kappa}_{1,1} & C^{\kappa,\kappa}_{1,2} & C^{\kappa,\kappa}_{1,3}\\
& & & & C^{\kappa,\kappa}_{2,2} & C^{\kappa,\kappa}_{2,3}\\
& & & & & C^{\kappa,\kappa}_{3,3} \\
\end{array} \right)_\ell
\label{Gell}
\end{equation}
where the upper diagonal block is galaxy clustering on its own, the lower diagonal block is tomographic weak lensing on its own, and the off-diagonal block is the cross-correlation between weak lensing and galaxy clustering. The data matrix is symmetric, and thus only the upper triangle is shown here. In Fig.~\ref{EuclidMatrixPlot} we depict the logarithm of such a full 10-bin matrix for a Euclid like survey.

The zeros in the off-diagonal block in our data matrices arise since shear-$a_{\ell m}$s must lie behind galaxy clustering $a_{\ell m}$s, in order to have a physically meaningful cross spectrum. 

The zeros in the galaxy clustering block arise from our non-overlapping redshift bin definition and since galaxy clustering is not an integral effect. Our redshift bin definition is depicted in Fig.~\ref{Redbins}, and uses a Euclid-like number density of galaxies of $30$ galaxies per $\rm{ arcmin}^2$, and a redshift dependency of \citep{Euclid,Stefano}
\begin{equation}
    n(z) = \frac{3}{2} \frac{z^2}{z_0^3} \exp\left(- \left[ \frac{z}{z_0} \right]^{\frac{3}{2}} \right),
\end{equation}
which is normalized to unity. We use the Euclid-typical value $z_0 = 0.9/\sqrt{2}$ following \cite{Stefano}.

For our redshift bin definitions we deviate slightly from the default Gaussian bin with equal galaxy number per bin. We replace the Gaussian by a fusion between a tophat and a Gaussian, realised by changing the power of the Gaussian from 2 to 2$\alpha$
\begin{equation}
    s(z,z_m,\sigma_z) \propto \exp\left( -\half \left[\frac{z-z_m}{\sigma_z}\right]^{2\alpha} \right),
\end{equation}
where $z_m$ is the mean of the redshift bin, and $\sigma_z$ is a parameter describing its width (the standard deviation in the Gaussian case). The parameter $\alpha$ can only take integer values, and we use $\alpha = 13$. For $\alpha = 1$, a Gaussian redshift-bin ensues, and for $\alpha > 1$, the sides of the bin begin to steepen up, approaching a tophat for $\alpha \to \infty$. Our choice of $\alpha$ was determined by requiring steep but smooth redshift bins, which are non-overlapping\footnote{For overlapping redshift bins, the off-diagonal elements in the galaxy clustering block would simply be non-zero.}.

This extremely useful redshift bin definition interacts well with the necessary integrations over Bessel functions when computing spherical harmonic power spectra in CLASS. It also leads to a more richly structured data matrix per multipole $\ell$ as given in Eq.~(\ref{Gell}). The latter facilitates the numerical inversion of the matrices. In fact, for redshift bins with large overlap and perfectly equal galaxy numbers, we found that the data matrices per $\ell$-mode are close to singular, as then all power spectra have similar amplitudes, and can thus nearly be written as linear combinations of each other. For the sake of numerical stability our somewhat more richly structured matrices proved highly reliable and correspond to a rather advantageous change of the survey setup.

Eq.~(\ref{Gell}) refers to a 3-bin tomographic survey, and the corresponding matrix $\hat{\sfG}_\ell$ for a 10-bin survey is 20-dimensional per $\ell$-mode. The matrices are implemented as such in our analysis, but here not shown due to their size. Our binning in redshift for the Euclid-like survey is shown in Fig.~\ref{Redbins}, which follows \cite{Euclid}. The means and the parameters $\sigma$ for our redshift bins are given in Tab.~\ref{tab:red}. Our setup assumes that the same redshift binning was used for both weak lensing and galaxy clustering, but this could be generalised.

\begin{table}
\caption{Redshift bins for our mock-Euclid survey. The bins are steepened-up Gaussian with $\alpha = 13$, mean redshift of $z_c$ and width $\sigma_z$.}
\label{tab:red}
\begin{tabular}{lcr}
bin number & central redshift $z_c$ & $\sigma_z$ \\ 
\hline
1 & 0.21 & 0.23 \\
2 & 0.545 & 0.065 \\
3 & 0.685 & 0.055 \\
4 & 0.825 & 0.05 \\
5 & 0.95 & 0.05 \\
6 & 1.07 & 0.05 \\
7 & 1.205 & 0.06 \\
8 & 1.382 & 0.08 \\
9 & 1.64 & 0.13 \\
10 & 2.41 & 0.55 \\
\hline
\end{tabular}
\end{table}

The joint data matrix of all Euclid $\ell$-modes is then block diagonal
\begin{equation}
    \sfX = 
\arraycolsep=1.4pt\def\arraystretch{1.0}
\def\arraystretch{1.0}
\left( \begin{array}{c|c|c}
 \hat{\sfG}_{\ell = 100} & 0 & 0 \\
\hline
0 & \ \ \ddots \ \ & 0 \\
\hline
0 & 0 & \hat{\sfG}_{\ell = 3000} \\
\end{array} \right)
\end{equation}
where each $\hat{\sfG}_\ell$ block is a 20 by 20 matrix, due to observing two fields (shear and galaxy distribution) in ten tomographic redshift bins. 

This matrix could be vectorized, in order to yield the usual data-vector,
\begin{equation}
    \fatx = \mathrm{vec}(\sfX),
\end{equation}
where $\mathrm{vec}$ is a vectorization operator that runs over all non-redundant elements of the data matrix (i.e.~over the upper triangle). We refrain however from this vectorization, since the matrix representation of the data can in some sense be understood as a from of dimensionality reduction of the data: the matrices have a special, prescriptive structure, and can directly be analyzed with a matrix-variate Wishart-likelihood \citep{SH15, Hamimeche2, SHH18}, rather than a extremely high-dimensional multivariate likelihood. Using these matrices enables us in the upcoming sections to never invert a huge $10^6$-dimensional covariance matrix, but multiple thousand $20\times20$matrices instead.

\subsection{Setup of the posterior}
Having laid out the concept of organising the data in matrices per $\ell$-mode, we continue to derive the posterior to be sampled. We denote conditional statements with a vertical bar, joint distributions by commas, and general probability densities by curly $\calP$.

For our assumptions of Section~\ref{sec:likelihood}, it follows that the data matrices per $\ell$-mode will contain cosmic variance, and shape- and shot-noise. The cosmic variance causes that our data matrices follow Wishart distributions, and Gaussian shot- and shape-noise then adds in a subsequent step. We focus on cosmic variance first, and abbreviate it by $\mathrm{CV}$. The observable data on the sky are then $\hat{\sfG}_\ell = \sfG^{\rm SN}_\ell$, where SN abbreviates shape- or shot-noise. This needs to be distinguished from the not directly observable $\sfG^{\rm CV}_\ell$, which neglects shot- and shape-noise, and only includes cosmic variance (CV). 

There exist two definitions of the Wishart distribution in the literature, only one of which has the correct skewness as it applies to cosmology. The form which applies for our estimators from Eq.~\ref{estimators}, is \citep{SH15,SHH18}
\begin{equation}
\begin{aligned}
    \mathcal{W}(\hat{\sfG}^\mathrm{CV}_\ell| \bar{\sfG}^\mathrm{CV}_\ell/\nu,\nu, p) & =
         A \exp\left(  - \frac{\nu}{2} \mathrm{Tr}[(\bar{\sfG}^\mathrm{CV})^{-1}_\ell \hat{\sfG}^\mathrm{CV}_\ell]\right),
    \end{aligned}
    \label{20wish}
\end{equation}
with the function $A$ being
\begin{equation}
    A = \frac{|\hat{\sfG}^\mathrm{CV}_\ell|^{\frac{\nu-p-1}{2} }  }{ 2^{\frac{p\nu}{2}}  |\bar{\sfG}^\mathrm{CV}_\ell / \nu|^{\frac{\nu}{2}} \Gamma_p\left( \frac{\nu}{2}\right) },
\end{equation}
and where determinants of matrices are indicated by vertical bars, e.g. $|\hat{\sfG}_\ell|$ is the determinant of the matrix $\hat{\sfG}_\ell$. The dimension of the matrices is $p$, which is a priori 20 for the Euclid-like survey, but will be less after cleaning our data set in section~\ref{sec:cleaning}. The trace is written as $\mathrm{Tr}$, and $\Gamma_p$ is the p-dimensional Gamma-function.

Euclid-like surveys are not cosmic variance limited: shape noise affects their weak lensing measurements, and Poissonian shot noise affects their galaxy clustering (GC) measurements. For galaxy clustering measured from galaxies with density $\bar{n}$ per tomographic bin, the Poissonian shot noise is $1/\bar{n}$. For weak lensing (WL), the intrinsic shape diversity of galaxies produces a shape noise variance $\sigma_\epsilon$, where we use the typical value $\sigma_\epsilon = 0.25$.

We model shape- and shot-noise according to the standard approach, see e.g.~\cite{CosmoLike}, using Gaussian distributions for these noises, and their variances are then
\begin{equation}
    \Sigma_{ii}^2 = 
\begin{cases}
& (\sigma_\epsilon^2/\bar{n})^2 / F_\ell \hfill (\rm{WL}),\\
& (1/\bar{n})^2/ F_\ell \hfill (\rm{GC}), \\
& \sigma^2_\epsilon/\bar{n}^2 / F_\ell\ \ \ \ \ \hfill (\rm{GC} \times \rm{WL}),\\
\end{cases}
\end{equation}
where $F_\ell = (2\ell+1)f_{\rm sky}$.
We pool all these variances into a joint covariance matrix $\Sigma$, which is diagonal in the sense of Eq.~(A1) of \cite{CosmoLike}, due to the Dirac delta function $\delta_{\ell,\ell'}$ enforcing an identification of the $\ell$-modes. In this paper we assume that clustering and weak lensing power spectra are measured from the same galaxies. For a Euclid-like survey with predicted galaxy number densities of 30 per $\mathrm{arcmin}^2$, this then leads to $\bar{n} = 3$ per $\mathrm{arcmin}^2$ per tomographic bin (since our tomographic bins have equal numbers of galaxies).

For a single realisation of cosmic variance, the shape- or shot-noise affected power spectrum then scatter around it in a Gaussian matter, which we denote by $\calG(\GSN | \GCV, \Sigma)$, where $\calG$ is the Gaussian distribution.
The hierarchical model for the posterior of cosmological parameters from Euclid-like observables is then
\begin{equation}
\begin{aligned}
\calP(\both | \GSN) & =  \int \calP(\both,\GCV | \GSN) \ \mathd \GCV\\
& = \int \frac{\calG(\GSN | \GCV, \Sigma) \calW(\GCV| \sfG_\ell(\both)) \pi(\both)}{\pi(\GSN)} \ \mathd \GCV.
\end{aligned}
\label{BHM}
\end{equation}
In other words, Eq.~(\ref{BHM}) is Eq.~(15) of \cite{SHH18} which began to derive the non-Gaussian likelihood for cosmic shear analyses after \cite{SHInsuff} found indication for non-Gaussianity in these data. Here, the likelihood Eq.~(\ref{BHM}) is now extended to cross power spectra with galaxy clustering, and written as a Bayesian Hierarchical Model instead of as a forward model. A noteworthy difference to \cite{SHH18} is however, that we here use the standard approach for the degrees of freedom $\nu$ (Eq.~\ref{nu}) as they arise from a continuous, Gaussian field. This approximation for the degrees of freedom was found to be incompatible with actual weak lensing simulations in \cite{SHH18}, with the real weak lensing data distribution function being more skewed than one would expect from Eq.~(\ref{nu}).

Resolving the skewness issue relies on heavy full-sky simulations of weak lensing, which is a lengthy progress which is not yet completed. For the time being, we thus use the standard approach for $\nu$, bearing in mind that $\nu$ is well isolated in our likelihood and can quickly be replaced upon availability of the required simulations. This approach implies that the Gaussian limit of our compound likelihood is the standard approach of \cite{CosmoLike}.

In this paper, we implement the integral over cosmic variance in Eq.~(\ref{BHM}) by sampling from the Wishart distribution $\calW(\GCV| \sfG_\ell(\both))$.

\begin{table}
\caption{Setup of the joint data set. The cosmic shear spherical harmonic power spectrum is $\kappa\kappa$, GC denotes galaxy clustering, $\kappa g$ is the cross power spectrum of (tangential) shear and galaxy clustering. The values follow the Euclid Red Book, but we (for now) cut at $\ell_{\rm max} = 3000$ instead of the ultimately targeted 5000 (since our training excludes baryonic feedback for now). }
\label{tab:surveys}
\begin{tabular}{lcccr}
\hline
$C_\ell$ & $\ell_{\rm min}$ & $\ell_{\rm max}$ & $z$-bins (cross-bins)  & $f_{\rm sky}$ \\
\hline
$\kappa\kappa$ & 100 & 3000 & 10 (55)  & 0.35 \\
GC & 100 & 3000 & 10 (55)  & 0.35 \\
$\kappa g$ & 100 & 3000 & 10 (55)  & 0.35 \\
\hline
\end{tabular}
\end{table}

Since our data matrix $\sfX$ is block-diagonal, the joint posterior for all $\ell$-modes is simply the product over the posteriors per $\ell$-mode. Tab.~\ref{tab:surveys} lists our cuts in $\ell$-range, together with the sky fraction which scales the degrees of freedom.

We therefore arrive at the posterior of parameters jointly inferred from cosmic shear, galaxy clustering and their cross-correlation for the Euclid-like survey,
\begin{equation}
\begin{aligned}
  \calP(\both|\sfX) \propto
  \prod_{\ell = 100}^{3000}\left[ \calP(\both | \GSN) \right]
 \prod_{i = 1}^{r}\left[ \pi(\theta_i)\right].
 \end{aligned}
\label{Posterior}
\end{equation}
Here, $r$ is the number of parameters to be inferred, and is only needed to loop over the priors on the parameters, given in Table~\ref{tab:example}. This is the posterior to be sampled for parameter inference. 

\subsection{Theoretical predictions for the power spectra}
In order to infer cosmological parameters, we require theoretical predictions for the spherical harmonics power spectra, for which we use CLASS with halofit \citep{Julien1801,CLASS1,CLASS2}. It is these theoretical power spectra, that we train our neural nets on: Section~\ref{sec:NN} will detail how we provide the cosmological parameters $\both$ as input to the nets, and train the nets such that they output the required power spectra.

\section{Neural Nets as content-addressable Memory}
\label{sec:NN}
Before using artificial neural nets to accelerate the computation of cosmological posteriors, let us shortly discuss why our use of neural nets still leaves perfect control over theory: the nets in our configuration do not `learn' anything new. In fact, the nets never analyze the \emph{data}, but simply `memorize' expensive \emph{theory} predictions. Our use of neural nets is therefore somewhat non-standard, as we do not distill information out of noisy data, but rather memorize a classical, noise-free function that varies over a wide parameter space. Our neural nets can hence not fit to noise, as there is none, which already removes one often voiced caveat against neural nets. The second caveat, loss of accuracy due to approximating, is dealt with by data cleaning in section~\ref{sec:cleaning}.

\begin{figure*}
\includegraphics[width=0.98\textwidth]{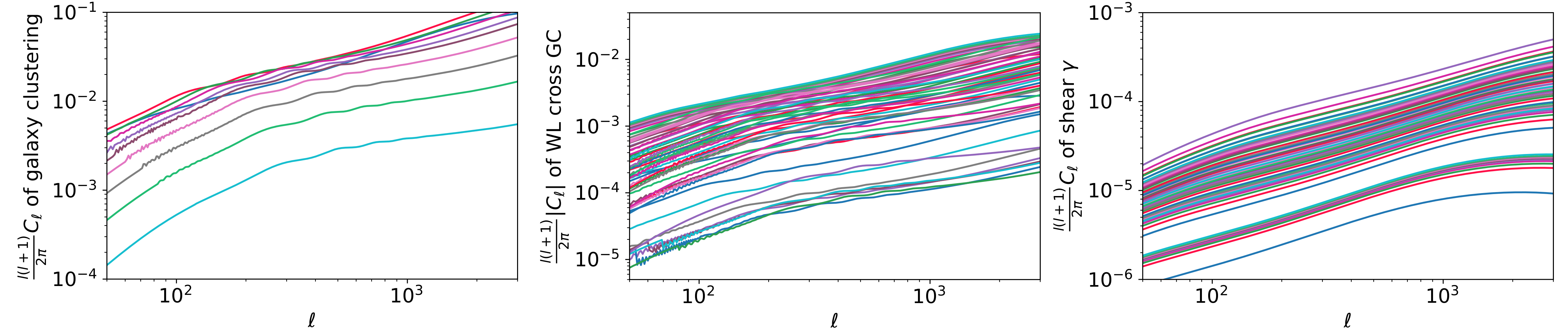} 
\caption{Theory predictions for the power spectra of our Euclid-like survey computed with CLASS. Plotted is a single cosmology, where the multitude of lines arises from the many redshift cross bins. From left to right we plot galaxy clustering, the cross power spectra between lensing and clustering, and weak lensing. Of each spectrum, the modes of $\ell \in [100,3000]$ are used in the likelihood, leading for our 120 spectra to a total of 348000 data points.  The jitter at low multipoles $\ell$ is numerical noise from the CLASS integration routines, and is also learned by the neural nets (but then removed from the likelihood, see section~\ref{sec:cleaning}).}
\label{Euclid_singleCosmology}
\end{figure*}

\begin{figure*}
\includegraphics[width=0.98\textwidth]{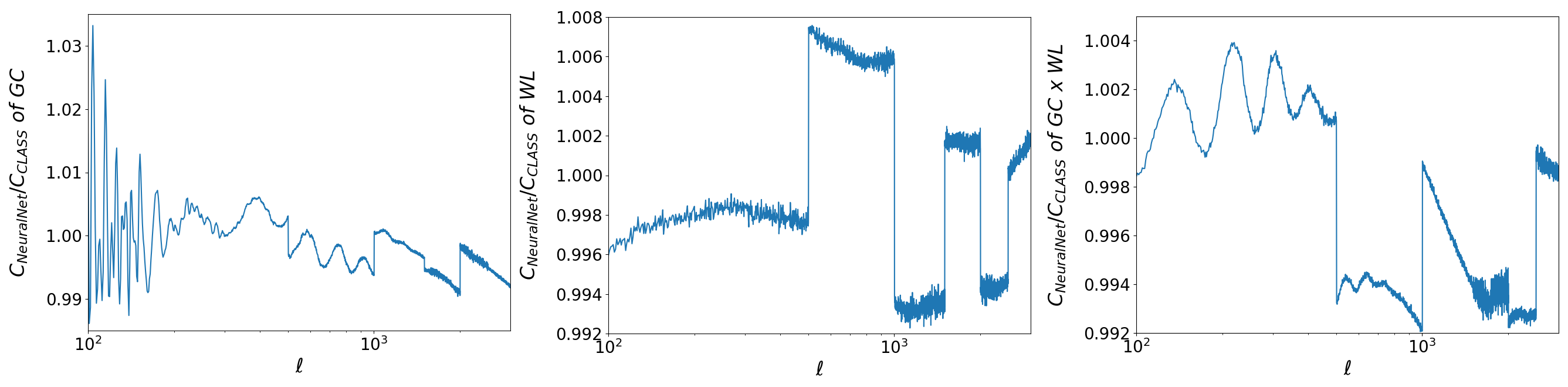} 
\caption{Achieved training accuracy for a random draw from the validation set, for all three power spectra types. Plotted is the ratio of the original power spectra computed with CLASS and the output power spectra of the fully trained nets, as a function of multipole $\ell$. The different segments in $\ell$-range correspond to the 6 neural nets which each coped with 500 $\ell$-modes. The total training accuracy is primarily below the sub-percent regime, apart from fitting to the baryonic acoustic oscillations in galaxy clustering spectra. The remaining inaccuracies are taken care of before using the nets in an MCMC sampler (see section~\ref{sec:cleaning}).}
\label{TrainingAccuracy}
\end{figure*}

\subsection{Training}
\label{training}
We train multiple neural nets to output the power spectra of weak lensing, galaxy clustering and their cross spectrum, as a function of seven cosmological parameters. The original CLASS computations for such power spectra are depicted in Fig.~\ref{Euclid_singleCosmology}, where each panel depicts the spectra for all redshift bin combinations. At Euclid-like precision, the neural nets need to achieve accuracies well below the sub-percent level. 

Our nets trained on 5238 power spectra computed with CLASS, over a seven-dimensional parameter space spanned by $\Omega_{\rm m}$ (the dark matter density), $\Omega_b$ (baryon density), h (Hubble factor), $\sigma_8$ (normalization of the power spectrum), $n_s$ (the spectral index of initial power spectra), $w_0$ (dark energy equation of state parameter), $w_a$ (evolution parameter of dark energy). The neural nets trained over the entire prior range given in Tab.~\ref{tab:example}.

We trained $3\times 6$ neural nets, where 6 nets trained on 500 distinct $\ell$-modes for one of the three power spectra types. The total range of $\ell \in [2,3000]$ was divided in 6 sub-ranges such that the size of the neural nets' output layers could be reduced from 3000 to 500, which in turn reduces the total number of free parameters in the neural nets. All our nets use three densely connected layers, with the input layer being 7-dimensional (corresponding to the cosmological parameters), followed by two hidden layers of dimension 128 and 256, followed by a 500 dimensional output layer corresponding to the trained $C_\ell$-predictions. Each hidden layer was followed by a dropout-layer with a 10-percent dropout rate, in order to stabilize training. 

During training, a major advantage was that our nets did not train on noisy data, but on classical functions. This reduces the total number of required training data as noise did not need to be suppressed. We observed a rapid increase in the training accuracy once more than 3000 training sets were passed for training.

This threshold can intuitively be understood: as the nets had to predict about 3000 $\ell$-modes per spectrum, degeneracies in training must be strong if less than 3000 training samples are provided. Once providing more than 3000 training samples, a good choice of architecture will become crucial for accurate training. This led us to sequentially shrinking our nets to ever smaller configurations, until arriving at the above described setup of $3\times6$ neural nets. During our iteration towards finding a good architecture, 20 percent of the total training set were left out of training and were instead separately used for validation.

After having settled on the final architectures, we iteratively generated four times 200 further Euclid-like predictions randomly across parameter regions where the neural nets showed poor accuracies in validation. Retraining on the additional 800 samples quickly increased the accuracy throughout the entire prior range. We then revalidated on further 136 validation sets, to assure our iterative search of good architectures did not lead to implicit over-fitting. 

We depict the final performance of the neural nets in Fig.~\ref{TrainingAccuracy}, where it is seen that the goal of sub-percent accuracy was indeed reached, with the exception of the baryonic acoustic oscillations seen as little wiggles around $\ell \approx 100$ in the galaxy clustering spectra. Further detail on the accuracy achieved throughout the entire prior range is given in appendix \ref{app:valid}, from which is can be seen that the accuracy is nearly constant through the prior range.

As a final note of caution, we point out that our public nets are trained for the redshift bins of Fig.~\ref{Redbins}. If the redshift bins are changed, the nets need to be retrained, just as they would need to be retrained for new theories.

\section{Data Cleaning: What did the neural nets not learn?}
\label{sec:cleaning}
Neural nets are by construction approximators, meaning that even after excessive training it can a priori not be expected that the neural nets produce the required functions perfectly, especially not for values of the cosmological parameters that they were not trained on. Our inference algorithm would thus be incomplete if we did not account for this loss of accuracy, which -- if disregarded -- would lead to biases in the inference.

As extended training will improve the accuracy of the neural net predictions, we chose to remove from the data any pattern that the neural net in its current training state does not resolve. To do so, we use differentiatability of physics, and positive-definiteness of our data- and theory-matrices: As the power spectra are differentiatable functions of the cosmological parameters, small changes in the cosmological parameters must lead to small changes in the power spectra. This results in a smooth variation of the likelihood as function of parameters. Any discontinuous changes of the likelihood values found during sampling with small step sizes are thus tell-tale signs that the networks did not fully capture the structure of the power spectra.

Secondly, we arranged our data set in data matrices per $\ell$-mode, according to Eq.~(\ref{Gell}), of which it is known that they must be the covariance matrix of $a_{\ell m}$ modes. At each $\ell$, all matrices must therefore be positive definite. If they are not, then this indicates inaccuracies of the neural nets. Crucially, such inaccuracies should not be heuristically `fixed' after training, as they are an expected outcome of the approximation, and we therefore impede these remaining inaccuracies from propagating into the analysis.

To do so, we let the neural nets compute multiple theory matrices $\sfG_\ell(\both_i)$ for multiple cosmological parameters $\both_i$. All matrices $\sfG_\ell(\both_i)$ are then diagonalized, and the eigenvalues are sorted by size. If the neural nets predict non-positive definite matrices, then negative eigenvalues will appear, and if the neural nets did not capture fine structures in the matrices, but only coarse overall structures, then the smallest eigenvalues will additionally be unstable. Therefore, removing all negative eigenvalues, and the smallest unstable positive eigenvalues, will remove the inaccuracy of the neural nets. Removing an eigenvalue then automatically necessitates the reduction of dimension, as the matrices would otherwise not be of full rank anymore.

It is however not permissible to arbitrarily remove a different number of eigenvalues at each point in parameter space, as this would correspond to explaining more or less data points for different parameter values. Rather, we construct ourselves a transformation that removes the unstable or not understood structures from the \emph{data}, and only the cleaned data set is then contrasted with the corresponding theory matrices. This treats all points in parameter space equally, and ensures the nets cannot fit inaccuracies to the data.

To implement this data cleaning algorithm, we use the following two properties of the Wishart distribution.

Firstly, Wishart distributions allow for congruent matrix transformations, in the sense of if $\sfG \sim \mathcal{W}(\boSig,n)$, then 
\begin{equation}
\sfC^{-1}\sfG\sfC^{-1,T} \sim \mathcal{W}(\sfC^{-1} \boSig \sfC^{-1,T},n).
\end{equation}

Secondly, Wishart distributions allow for dimensionality reduction if the matrices are partitioned. Let the $p\times p$ matrices $\sfG$ and $\boSig$ be partitioned in the same sub-blocks
\begin{equation}
    \sfG = \begin{pmatrix}
    \sfG_{11} & \sfG_{12}\\
    \sfG_{21} & \sfG_{22}
    \end{pmatrix}, \ \ \ \ \
    \boSig = \begin{pmatrix}
    \boSig_{11} & \boSig_{12} \\
    \boSig_{21} & \boSig_{22}
    \end{pmatrix},
\end{equation}
where $\sfG_{11}$ and $\boSig_{11}$ are $q\times q$ matrices with $q < p$. Then it follows from $\sfG \sim \mathcal{W}(\boSig,n)$, that $\sfG_{11} \sim \mathcal{W}(\boSig_{11},n)$. To remove negative or unstable eigenvalues through dimensionality reduction, the Wishart distribution therefore allows us to first diagonalize, and then determine a new dimension $q<p$ at will, as long as $q$ is then kept fixed when sampling the posterior. The latter implies that all points in parameter space are analyzed with the same likelihood function, which uses the same -- cleaned-- data set at each point.

The initial dimension of our matrices is $p = 20$. Before looking at the data, we thus compute multiple theory matrices $\sfG(\both_i)$, and diagonalize these
\begin{equation}
    \sfG(\both_i) = \sfC(\both_i) \mathrm{diag}(g_{1,1},...,g_{20,20})\sfC(\both_i)^T.
    \label{trafo}
\end{equation}
The basis changing matrices $\sfC(\both)$ depend on cosmological parameters, as the theory matrices cannot be expected to be co-diagonal for all parameter values\footnote{This, and the change of eigenvalues as a function of parameters, is what the Wishart likelihood reacts to when inferring parameters.}. The index $q$ is then picked to discard negative or unstable eigenvalues, starting at the smallest. We found that $q = 16$ reliably removes all negative eigenvalues, and $q = 15$ removes the smallest unstable eigenvalues whereupon the likelihood becomes smooth. To be on the safe side, we cut at $q = 13$ which discards two more of the smallest eigenvalues. 

Together with $q = 13$, we pick one matrix $\sfC(\both_\mathrm{o}) \equiv \sfC $ per $\ell$. The chosen $\both_\mathrm{o}$ is arbitrary, because $\sfC(\both_\mathrm{o})$  must be kept fixed  when sampling the reduced Wishart distributions. In summary, our parameter inference replaces the original 20-dimensional Wishart likelihood $\mathcal{W}(\hat{\sfG}^\mathrm{CV}_\ell| \bar{\sfG}^\mathrm{CV}_\ell/\nu,\nu, p)$ of Eq.~(\ref{20wish}) by the 13-dimensional 
\begin{equation}
\mathcal{W}\left(\sfC^{-1}_\ell \hat{\sfG}^\mathrm{CV}_\ell \sfC^{-1,T}_\ell| \frac{1}{\nu}\sfC^{-1}_\ell \bar{\sfG}^\mathrm{CV}_\ell \sfC^{-1,T}_\ell,\nu, q\right).
\end{equation}
The transformation $\sfC^{-1}_\ell \bar{\sfG}^\mathrm{CV}_\ell \sfC^{-1,T}_\ell$ linearly superimposes the different power spectra per $\ell$-mode, where the superposition coefficients are products of the elements of the matrix $\sfC^{-1}_\ell$. We hence compute the same superposition for the shape- and shotnoise and this provides the basis for the posteriors shown in section~\ref{sec:MCMC}. Note, that in comparison to \cite{Moped}, our dimensional reduction here is \emph{not} a lossless compression of the data: some constraining power is indeed lost due to removing the nets' inaccuracies, and can only be fully captured by extended training.

\section{MCMC forecasts for a Euclid-like survey}
\label{sec:MCMC}

\begin{figure*}
\includegraphics[width=0.98\textwidth]{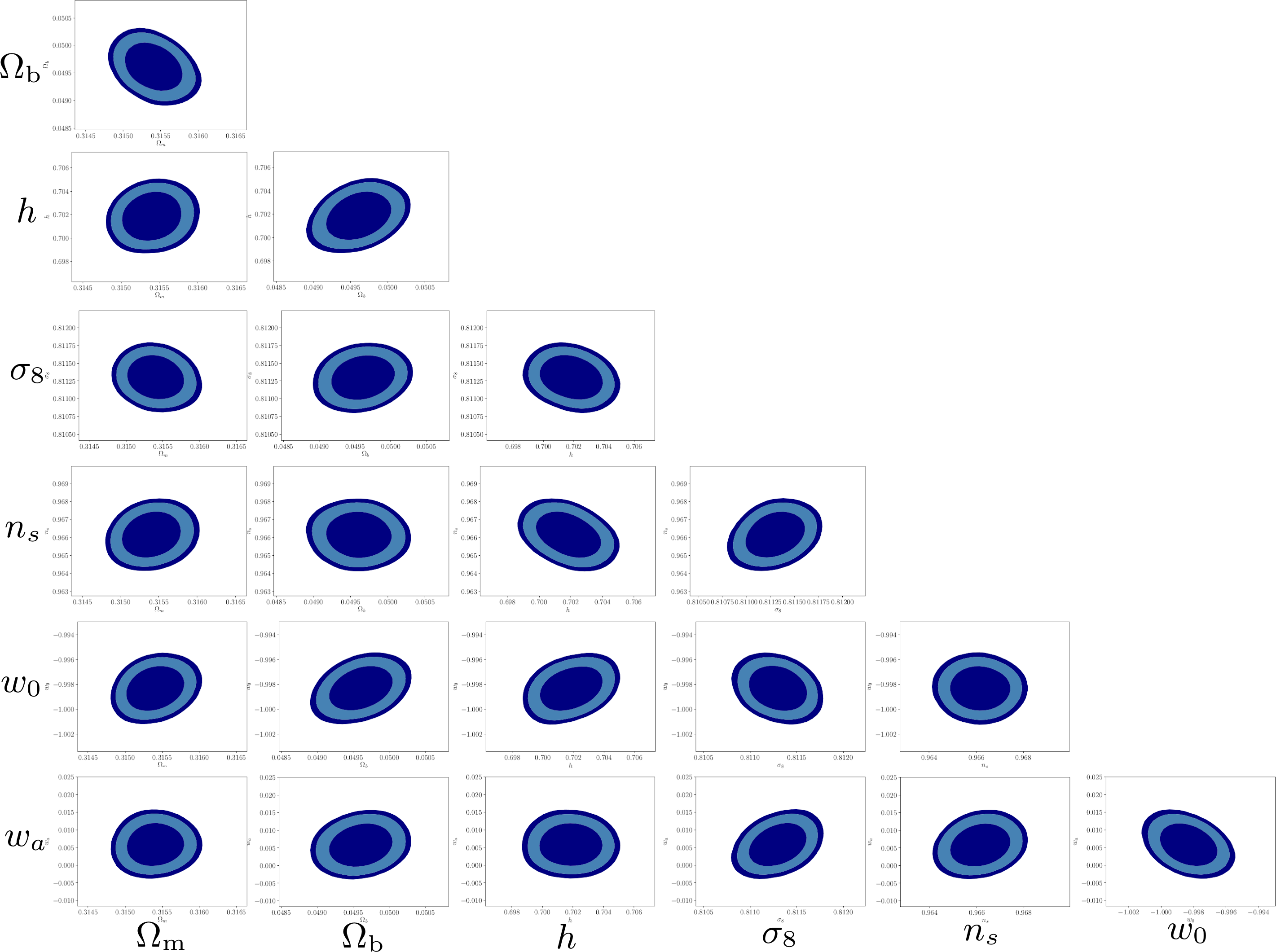} 
\caption{Forecasted marginal posterior contours for a $w$CDM model, using our synthetic data sets of a Euclid-like survey in a 3x2 setup, combining weak lensing, galaxy clustering and tangential shear measurements over an $\ell$ range from 100 to 3000. Cosmic variance is implemented as a Wishart distribution, shape- and shot-noise follow Gaussian distributions. The contours contain 68, 90 and 95 percent of posterior volume. All nuisance parameters for redshift uncertainties, baryonic feedback, galaxy bias, intrinsic alignments, etc have been omitted, hence the contours here shown are a proof of concept of sampling with calls to a neural net which memorised the seven primary parameters shown.}
\label{Forecasts}
\end{figure*}

To showcase our method, we run the algorithm to create posteriors of cosmological parameters for a simplified Euclid-like analysis. Fig.~\ref{Euclid_singleCosmology} plots the theoretically predicted power spectra for our Euclid-like survey. The there shown multitude of lines is for a single cosmology, and the multitude arises from the redshift binning only. We model the survey according to table~\ref{tab:surveys}; using each $\ell$-mode, we arrive at a total number of data points of 348000. We then create a synthetic noise-free data set for the fiducial cosmology
\begin{equation}
    \begin{aligned}
    & \Omega_{\rm m}  = 0.315,\ \Omega_{\rm b} = 0.0492,\ h = 0.7, \sigma_8 = 0.811,\\
    & n_s = 0.965,\ w_0 = -1,\ w_a = 0.
    \end{aligned}
\end{equation}
A future longterm-goal of the algorithm at hand is to free up computational resources for handling nuisance parameters related to redshifts, galaxy bias, etc, in a Bayesian hierarchical model. These nuisance parameters do not have the same significance as the primary cosmological parameters, would occur on a different level in a hierarchical likelihood than the fundamental theory, and are therefore omitted from the networks who only train on the fundamental parameters. This also implies that the posterior calculation spares out approximately 20 or more nuisance parameters, and the size of contours in Fig.~\ref{Forecasts} is not to be regarded as representative of a Euclid-like survey. A forecast with nuisance parameters is given in \cite{Julien1801}.

Instead, the posterior of Fig.~\ref{Forecasts} is a proof of concept, which showcases Metropolis-Hastings sampling that calls the trained neural nets to compute the theoretical power spectra, and which reduces the dimension to $q=13$ as described in Sec.~\ref{sec:cleaning}. Crucially, even though computing the training data and training the networks required multiple months of CPU time and intermittent use of GPU-based high performance computing facilities, the posteriors of Fig.~\ref{Forecasts} were computed on a usual desktop within a day.

\section{Discussion}
\label{sec:conclusions}
In this paper we have designed a joint algorithm of artificial neural nets and a Monte Carlo Markov sampler, in order to sample cosmological posteriors. Our aim was to provide an automatic acceleration of cosmological computations, as here enabled by the neural nets being used as a `memory' for expensive physical calculations. A vital step in our algorithm was to avoid that expected inaccuracies in the neural net approximations propagate into the inference of physical parameters where it could cause biases. We achieved this by cleaning the data set in order to remove fine structures which the neural nets did not capture.

We demonstrated the capabilities of the algorithm for a Euclid-like data set, analysed with a likelihood that omits (for now) all nuisance parameters.

Our nets here trained use a $w$CDM model, and new nets need to be trained for beyond-$w$CDM cosmologies. In the long run, these nets can be merged, and especially be made publicly available, where the latter will drastically cut computational needs for even further training and use.

Whether or not this will one day channel into a `universal net' to memorise theoretical physics is an open question. Of paramount importance is however, that in our algorithm, the neural net can always be switched off, and the inference then falls back onto traditional MCMC sampling. This means the net still leaves full freedom and control to the physicist, and theoretical understanding can progress as previously.

\section{Acknowledgements}
We thank Julien Lesgourgues and his group for maintaining CLASS at a superb standard. We thank our new colleague Simon Portegies Zwart and Leiden's computer group for sharing their computational infrastructure. Further inspiring discussions with Euclid's simulation working group is thankfully acknowledged. ES is supported by Leiden's Oort-Fellowship programme.

\appendix

\section{Training accuracy and validation}
\label{app:valid}
\begin{figure*}
\includegraphics[width=0.98\textwidth]{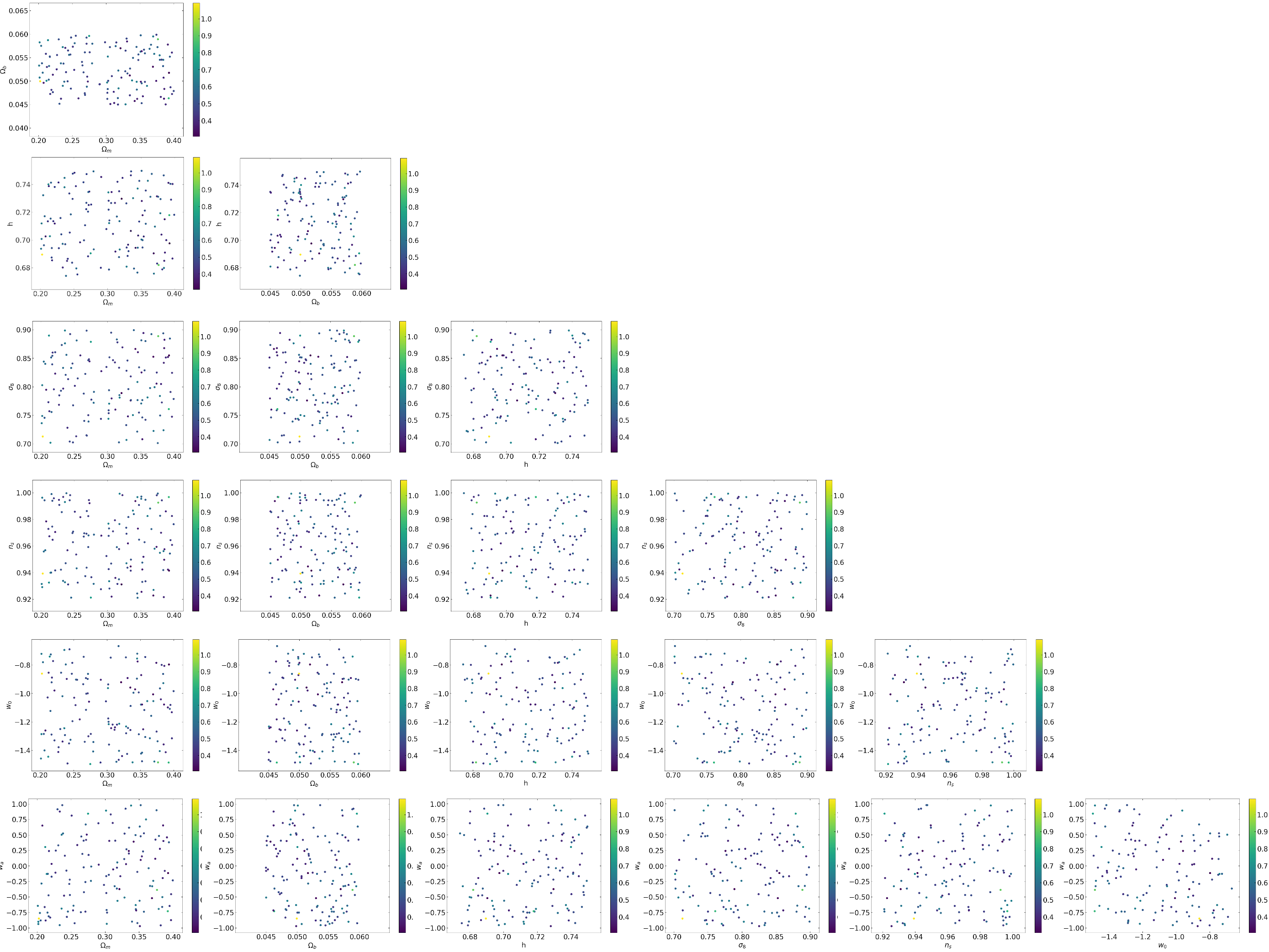} 
\caption{Achieved accuracy of the neural nets when predicting unseen validation sets. The colour bar indicates the accuracy averaged over all $\ell$-modes and averaged over all spectral types. The here depicted accuracy is defined in Eq.~(\ref{finalmre}).     }
\label{validationset}
\end{figure*}

After training the neural nets, their achieved accuracy was validated on an independent validation test set. As measure of total accuracy we defined the mean relative error
\begin{equation}
    \textrm{MRE} = \frac{1}{120}\sum_{i = 1}^3 \frac{1}{N_\ell}\sum_{\ell = 100}^{2999} \frac{C^{\textrm{CLASS}}_{\ell,i} - C^{\textrm{net}}_{\ell,i}}{C^{\textrm{CLASS}}_{\ell,i}},
    \label{finalmre}
\end{equation}
which averages over all $\ell$-modes, and over the 120 spectral types being the galaxy clustering power spectrum per redshift bin (10 spectra), the cosmic shear auto- and cross-spectra (55), and the tangential shear spectra (55). The final validation test set contained 136 full Euclid-like theory vectors, where each contained all 120 spectra. For each of these 136 validation sets the final mean relative error of Eq.~(\ref{finalmre}) is depicted in Fig.~\ref{validationset}.

Fig.~\ref{validationset} reveals that the trained nets achieved a sub-percent accuracy in predicting unseen Euclid-like theory spectra through the entire prior range of Tab.~\ref{tab:example}.

\bibliographystyle{mn2e}
\bibliography{TDist}

\label{lastpage} 
\bsp 
\end{document}